\documentclass[12pt]{article}
\usepackage{graphicx}
\newcommand{\beq}{\begin{equation}}
\newcommand{\eeq}{\end{equation}}
\newcommand{\bea}{\begin{eqnarray}}
\newcommand{\eea}{\end{eqnarray}}

\def\fun#1#2{\lower3.6pt\vbox{\baselineskip0pt\lineskip.9pt
\ialign{$\mathsurround=0pt#1\hfil##\hfil$\crcr#2\crcr\sim\crcr}}}
\begin{document}
\begin{titlepage}

\begin{flushright}
{\tt hep-ph/0512nnn}\\
FIT HE-05-05 \\
\end{flushright}
\vspace*{3mm}
\begin{center}
{\bf\Large Holographic Model for Mesons \\
\vspace*{3mm}
at Finite Temperature \\ }
\vspace*{12mm}
Kazuo Ghoroku\footnote[1]{\tt gouroku@dontaku.fit.ac.jp},
Masanobu Yahiro\footnote[4]{\tt yahiro2scp@mbox.nc.kyushu-u.ac.jp}
\vspace*{2mm}

\vspace*{7mm}

{\em ${}^{*}$Fukuoka Institute of Technology, 
Fukuoka 811-0295, Japan} \\
\vspace*{2mm}
{\em ${}^{\S}$Department of Physics, Kyushu University, 
Fukuoka 812-8581, Japan} 
\end{center}

\vspace*{20mm}

\begin{abstract}
A holographic model of QCD at finite temperature is proposed and 
thermal properties for light mesons are examined in terms of this model. 
We find interesting temperature dependences of masses and decay constants
for several mesons, and Nambu-Goldstone theorem for the chiral symmetry
breaking and generalized Gell-Mann-Oakes-Renner relation are assured
at finite temperature. Furthermore, we can see in the chiral
limit that the speed of the pion in the thermal medium decreases to zero
when the temperature approaches to the critical point of quark 
confinement and deconfinement transition, where the chiral symmetry is 
restored. 

\end{abstract}
\end{titlepage}

{\bf 1.} Recently, from the gravity/gauge correspondence, probe brane
approach to the gauge theory with flavor quarks
has been largely developed in terms of
the system of $D_p/D_{p+4}$ branes \cite{AdS,GSUY}. In these, flavor quarks
have been introduced by embedding $D_{p+4}$ branes as probe
in the 10d background,
which is deformed to the anti-de Sitter space (AdS) by the stacked 
$D_p$ branes. 
Then several important dynamical properties of QCD as chiral symmetry 
breaking and quark confinement have been explained, and furthermore, the
flavored-meson masses have been calculated.

Inspired by these works, phenomenological, 
five-dimensional holographic models have 
been proposed to explain more quantitatively the physical
properties of light mesons in a simpler setting \cite{EKSS,RP,TB,GMTY,KLS}.
Up to now, all these works of this framework are restricted to the case of 
zero temperature ($T=0$).
Meanwhile, it is interesting to extend this simple framework to the 
finite temperature scheme in order to compare the thermal properties of the 
quantities estimated in this framework and the recent experiments \cite{STAR}.
Up to now, such thermal properties of light mesons have been estimated by the 
lattice, chiral perturbations and sigma models. So, we extend here the model 
proposed at $T=0$ in \cite{EKSS,RP} to the finite-temperature version in
order to give 
the first holographic approach in this category.

In \cite{EKSS,RP},
the gravity dual of the gauge theory with confinement at $T=0$ is expressed
by the 5d AdS with an infrared (IR) cutoff at an appropriate position
of the fifth coordinate. 
This IR cutoff is essential to form a gravity dual of
a gauge theory in the confinement phase, and the position of IR cutoff is
very sensitive to the meson mass.
In this sense, this point is 
fixed rigidly.
The finite-temperature version is analogously obtained by 
replacing the AdS background with the AdS- Schwarzschild, and 
the horizon, which is related to the temperature of the gauge theory, 
in the AdS- Schwarzschild is set inside the
IR cutoff at low temperature, then the horizon is hidden 
by the IR cutoff in the 5d bulk. 

It is well-known that
the 5d AdS-Schwarzschild background corresponds to the finite
temperature version of CFT with deconfinement phase
when the IR cutoff is removed. This situation is realized in the present case
when the IR cutoff is put inside of the horizon. In this case, the 
IR cutoff is covered by the horizon, then the effect of the
IR cutoff disappears. This setting corresponds to the high temperature phase
of our model. On the other hand, at low temperature phase,
the cutoff is set outside, the fields in the bulk
end at this IR cutoff
and appropriate boundary conditions for the fields are
imposed at this end point. As a result, 
we find infinite series of discrete
meson spectra. This situation is completely parallel to the AdS
case, but the system is set at finite temperature in the present case.

The temperature is understood from the finiteness of the Euclidean time
in the AdS-Schwarzschild configuration in order to evade the 
conical singularity which may arise at
the horizon. This constraint for the Euclidean time should
be preserved, even if the IR cutoff is introduced, in order to connect 
the manifold smoothly up to the critical temperature of quark confinement
and deconfinement ($T_c$).
So we can perform the analysis for the properties of the light mesons
in the region of $0<T<T_c$, 
by this configuration.
As for the quark part, we use the effective 5d action 
($S_{\rm meson}$) introduced
in \cite{EKSS,RP} for the sake of the simplicity.

\vspace{.3cm}
{\bf 2.} The bulk background, which is dual to the finite temperature 4d QCD,
is formed by
introducing the IR cutoff \cite{EKSS,GMTY} $z_m$ into the 5d AdS-Schwarzschild 
solution,
\bea
   ds^2_5={1\over z^2}\left(-f^2(z)dt^2+(dx^{i})^2+{dz^2\over f^2(z)} \right)
      \ , \quad f^2(z)=1-({z\over z_T})^4 \ . \label{fmet}
\eea
where the radius of AdS${}_5$ is taken as unit. As is well-known,
the imaginary time is restricted as $0\leq it=\tau <\pi z_T$,
to avoid the conical singularity. Then 
the temperature $T$ is given as $T=1/(\pi z_T)$.
As for the infrared cutoff $z_m$,
we consider the case of $z_m<z_T$ in order to see the confinement phase
at finite temperature. In this case, fields are restricted to the region
$0\leq z \leq z_m$. 

The 
condition 
$0\leq it=\tau <\pi z_T$ is not altered even if the cutoff is introduced.
Then the temperature is retained as given above.
When the cutoff has been introduced, however, the possible conical point is
removed for $z_m<z_T$, so it may be possible to change the temperature
in order to find a smooth manifold for $z<z_m$. 
But here we fix the temperature
as $T=1/(\pi z_T)$ in order to extend smoothly
the region of $z_T$ up to $z_T=z_m$.
The confinement is thus assured since we have
infinite series of discrete meson spectrum, quark and anti-quark bound states,
due to the boundary at $z_m$. 

On the other hand, for the case of $z_m>z_T$, the situation is the same with 
the finite temperature CFT. In this case, 
the cutoff has no meaning since $z$ is restricted to $z_T>z>0$.
Then we find the deconfinement phase with a few number of possible meson 
states, whose masses should be below the double quark-mass. This point 
is assured in \cite{GSUY}. 
This implies that the critical temperature $T_c$ of 
confinement-deconfinement is set in the present case
at $T_c={1/(\pi z_m)}$ which is estimated here
to be $\sim 102 $~MeV from meson spectra. This value is not so wrong since 
our model is not so strict.

\vspace{.3cm}
{\bf 3.} For the sake of the simplicity,
according to \cite{EKSS,GMTY}, we start with
the following 5d meson-action under the background 
(\ref{fmet}). This action is considered as the quark part represented
by the probe brane in the string theory. And the fields on the brane
represent meson states regarded as the bound states of quark and 
anti-quark.
The fields considered here are the gauge fields, 
$L_M$ and $R_M$, and a scalar field $\Phi$ whose vacuum expectation value
(VEV) is connected to chiral symmetry breaking. Here, 
we consider ${N_f=2}$ flavors, and $\Phi$ transforms 
as a $({\bf 2_L}, {\bf 2_R})$. The action is
\begin{equation}
S_{\rm meson}=\int d^4x dz\, \sqrt{-g}\, {\rm Tr}\left[-\frac{1}{4g_5^2} ({L_{MN}L^{MN}}
+{R_{MN}R^{MN}}) - {|D_M\Phi|^2} - M^2_\Phi|\Phi|^2\right]\, ,
\label{3-1}
\end{equation}
where the covariant derivative is defined as 
$D_M\Phi=\partial_M \Phi+iL_M\Phi-i\Phi R_M\,$, 
$g$ is the determinant of the metric, and $L_M=L_M^a\tau^a$ for 
the Pauli matrices $\tau^a$ and
similarly for other fields. $g_5$ is the 5d gauge coupling.
We define $\Phi=S\, e^{i\pi^a\tau^a}$ 
and ${1\over 2}v(z)\equiv \langle S \rangle$, where
 $S$ corresponds to a real scalar and $\pi$ to a real pseudoscalar
($S\rightarrow S$ and $\pi\rightarrow -\pi$ under  $L\leftrightarrow R$).
They transform  as  ${\bf 1+3}$ under SU(2)$_V$.

\vspace{.3cm}
{\bf 3-1.} The 5d mass of the scalar is set as
$M_{\Phi}^2 =-3$ to consider the bulk field which corresponds in the
gauge theory side to an operator with the conformal dimension $\Delta=3$. 
Then the equation of
motion for $v$ is given as,
\bea
\left[ \partial_z^2 - \frac{4-f^2}{zf^2} 
\partial_z + {3\over z^2f^2}\right]v(z)=0. 
\label{mscalar}
\eea
And we obtain 
\beq
 v(z)={z}\left({m_q}~{}_2F_1({1\over 4},{1\over 4},{1\over 2},{z^4\over z_{T}^4})
  +c~z^2~{}_2F_1({3\over 4},{3\over 4},{3\over 2},{z^4\over z_{T}^4})\right).
\label{profile}
\eeq
Here two integral constants,
$m_q$ and $c$, are identified with the quark mass (explicit
breaking of the chiral symmetry) and the chiral condensate 
(spontaneous breaking of chiral symmetry in the chiral limit), respectively. Actually,
we get the approximate form, $v(z)\sim m_qz+cz^3$ near $z=0$,
which is consistent with the above statement.

We notice here
that $v(z)$ varies with temperature $T$ or $z_T$, which is specific to
our model, but the parameters $m_q$ and $c$ are independent
of $T$. In other words, the gauge theory 
considered here is characterized by these three parameters.
{However, the chiral condensate $c$ should be determined dynamically
and it should depend on $T$. This point is the defect of the present model,
and we must be careful about the results of our calculations for the
quantities, which depend on this "parameter" $c$, when we discuss the
$T$ dependence of them.}

\vspace{.3cm}
The fluctuation of 
$S$, which is defined as $S=v(z)/2+\sigma$, can be
observed as a singlet meson state ($\sigma$). 
{ Here and hereafter we consider
the static mode, $\partial_i\phi=0$, for any field $\phi$ in order to
derive the mass in a simple way. So the invariant
mass, or pole mass, is defined here as $-\partial_{t}^2\phi=m^2\phi$.} Then
the equation for $\sigma$
is given by adding this 4d mass term to Eq.~(\ref{mscalar}) as
\bea
\left[ {m^2\over f^4} + \partial_z^2 - \frac{4-f^2}{zf^2} 
\partial_z+{3\over z^2f^2} \right]\sigma = 0. 
\label{scalareom}
\eea
\\
It should be noticed that
this equation is independent of $v(z)$. 
In this sense, the model given here is insufficient since all the fluctuations
on the probe brane would be influenced by the shape of the embedded brane.
But the improvement 
of this defect is postponed to the future.

The discrete mass-spectrum 
is obtained by solving this equation with the boundary conditions,
$\sigma(z) |_{z_0} = \partial_z \sigma(z) |_{z_m} = 0$, where $z_0$
is the UV cutoff which is taken to zero after all. The mass depends on $z_m$
and $T$. By using the value of $z_m$ determined by the
$\rho$ meson mass at $T=0$ as $1/z_m=.323$~GeV \cite{GMTY}, 
the $T$-dependence of the $\sigma$ meson mass
is estimated. The numerical results are shown in the Fig.\ref{vector}. 
We can observe that the mass decreases with 
increasing $T$ and approaches to zero for 
$T\to T_c$. This behavior is
easily understood from Eq.(\ref{scalareom}). The mass term, ${m^2/f^4}$, 
in the equation is largely
enhanced near $z=z_T$, where $f\to 0$,
so $m$ should be very small when
$z$ approaches to $z_T$. This situation is realized for $z_T\to z_m$
or $T\to T_c$. This point is therefore the most prominent check point of this
model. The experiments to assure this property are acquired.

\vspace{.3cm}
{\bf 3-2.} The gauge bosons are separated to
the vector and the axial vector bosons $V_M$ and $A_M$, and are defined as 
$L_M \equiv V_M + A_M$ and $R_M \equiv V_M - A_M,$ respectively. 

First of all, we consider the vector mesons. At finite temperature, it is
impossible to write the equation of motion covariantly, so we consider
the linearlized equation for the spatial component $V_{i}$ here.
And it is (employing $V_{z}=0$ gauge) given as
\beq
\left[ {m^2\over f^4} + \partial_z^2 - \frac{4-3f^2}{zf^2} 
\partial_z \right]V_i=0.
\label{vectoreom}
\eeq
This equation gives us the discrete eigenvalues $m^2=m_n^2$.
{In order to see the decay constants, it is convenient to expand
$V_i(x,z)$ as $V_i(x,z)=\sum_{n}V_i^{(n)}(x) h_n^V(z)$ for each
mass eigen-state. The spacial components of the
wave-function for each mode are normalized as
\beq
 \int^{z_m}_{z_0}dz {(h^V_n(z))^2\over z~f^2}=1, \label{normalization}
\eeq
where we notice the factor $1/f^2(z)$ in this integrand. This is
an important factor in determining the $T$ dependence of the decay
constant near $T=T_c$. Actually, we find rapid decreasing of $F_{\rho}$
near $T=T_c$ due to this factor as shown in Fig.\ref{vector}.

The mass of vector meson is given by
the boundary condition 
$V_i(z_0) = \partial_z V_i(z_m) = 0$ 
similarly to the
case of the sigma meson. 
Note that Eq.(\ref{vectoreom}) is also independent of $v(z)$, then
the masses of vector meson depend only on $T$, and the $T$-dependence
of the lowest one, which is identified with $\rho$ meson, 
is shown in Fig.\ref{vector}.

\begin{figure}[htbp]
\begin{center}
\voffset=15cm
  \includegraphics[width=6cm,height=5cm]{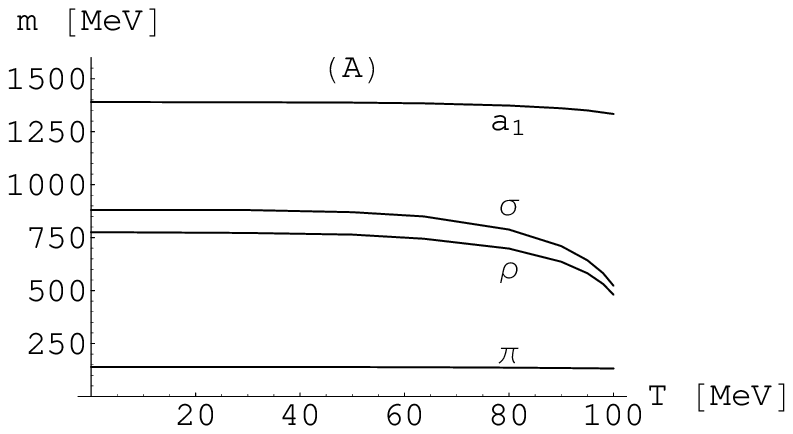}
  \includegraphics[width=6cm,height=5cm]{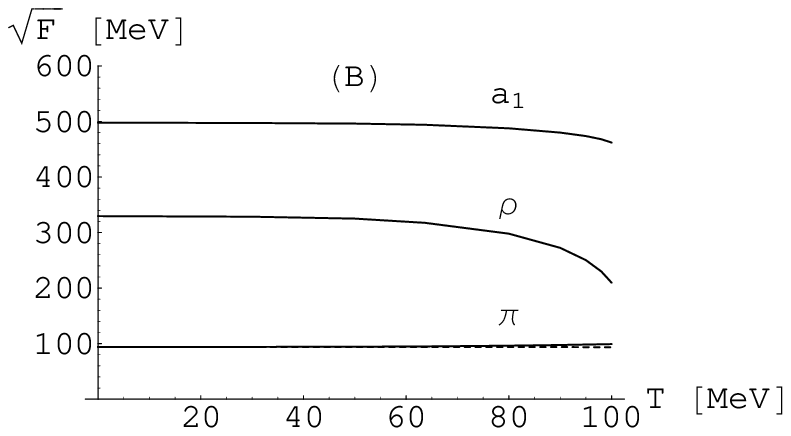}
\caption{ (A) The curves show the $T$ dependence of $m_{\pi}$, $m_{\rho}$, 
$m_{\sigma}$ and $m_{a_1}$ from the bottom. In (B), the solid curves denote 
the $T$ dependence of $F^t_{\pi}$, $\sqrt{F_{\rho}}$ and $\sqrt{F_{a_1}}$ from 
the bottom. The dotted curve denoting $F^s_{\pi}$ is constant 
with high accuracy at $T< 100$~MeV. 
}
\label{vector}
\end{center}
\end{figure}

We can see that the mass of $\rho$ meson vanishes or becomes
very small near $T=T_c$. This is consistent with
the expectation obtained from the equation (\ref{vectoreom}) as in 
the case of the $\sigma$-meson.

\vspace{.3cm}
{\bf 3-3.} The linearized equations of motion 
for the axial vector meson $A_\mu$ and the pion $\pi$ are obtained 
by decomposing $A_\mu$ into the transverse and 
the longitudinal part, $A_\mu = A_{\mu \perp} + \partial_\mu \varphi$; 
for simplicity, flavor index is neglected. 
Using $A_{z}=0$ gauge, one then one obtain   
\bea
\label{axial1}
&&[
{m_a^2\over f^4} + \partial_z^2 - \frac{4-3f^2}{zf^2} 
\partial_z - g_5^2{v^2\over z^2f^2}] A_{i \perp} = 0, 
\\ 
\label{axial0}
&&[
\partial_z^2 - \frac{1}{z} \partial_z 
- g_5^2{v^2\over z^2f^2}] A_{0 \perp} = 0, 
\\
\label{axial2}
&& [\partial_z^2 - \frac{1}{z}\partial_z]\varphi 
- g_5^2 {v^2\over z^2f^2} (\pi + \varphi) = 0, 
\\
\label{axial3}
&&
m_{\pi}^2\partial_z \varphi+ {g_5^2 {v^2 f^2 \over z^2}}\partial_z \pi=0,  
\eea
where the coupling constant $g_5$ is determined from the vector current 
two-point function at $T=0$~\cite{EKSS}. 
Here $m_{a}$ and $m_{\pi}$ are the pole masses defined as 
$-\partial_{t}^2\varphi=m_{a}^2\varphi$ and 
$-\partial_{t}^2\pi=m_{\pi}^2\pi$. 
These equations are solved numerically under the boundary conditions, 
$A_{\mu \perp}(z_0)=\partial_zA_{\mu \perp}(z_m)=0$ and 
$\varphi(z_0)=\partial_z\varphi(z_m)=\pi(z_0)=0$.

\vspace{.3cm}
The decay constants of the axial mesons and the pion are 
calculated from the wave functions as~\cite{EKSS} 
\beq
F_{a_n}^2 = \frac{1}{g_5^2} \left[ 
\left. \frac{d^2h^{A}_n}{dz^2} \right|_{z_0} \right]^2~(n \ne 0), \quad
({F^{t,s}_\pi})^2 = - \frac{1}{g_5^2} 
  \left. \frac{\partial_z {A}^{(0)}_{0,i \perp}}{z} \right|_{z_0} ,  
\label{F-pi}
\eeq
where 
$A_{i \perp}(x,z) = \sum_{n} \alpha^{(n)}_{i}(x) h_n^A(z)$ 
and 
$h_n^A(z)$ is normalized as $h_n^V(z)$ given in (\ref{normalization}).
Furthermore, $A^{(0)}_{0 \perp}$ ($A^{(0)}_{i \perp}$) is 
the zero-mode solution to 
Eq. (\ref{axial0}) (Eq. (\ref{axial1}) with $m_a^2=0$), 
satisfying 
$A^{(0)}_{0,i \perp}(z_0)=1$ and $\partial_z A^{(0)}_{0,i \perp}(z_m)=0$, 
and $F^{t,s}_\pi$ are the timelike and spatial components 
of the pion decay constant. Obviously, 
$F^{t}_\pi$ and $F^{s}_\pi$ are different from each other at finite $T$, 
while $F^t_\pi=F^s_\pi$ at $T=0$.

The masses and the decay constants of axial-vector mesons and pion 
depend on four parameters $m_q$, $c$, $z_m$ and $T$ through $v(z)$ and $f(z)$ 
in Eqs. (\ref{axial1})-(\ref{axial3}), 
while those of vector mesons do only on $z_m$ and $T$. 
For the consistency between the vector and axial-vector meson sectors, here 
we take the same $z_m$ as that determined in the vector meson sector; 
namely $1/z_m=0.323$~GeV. Parameters, $m_q$ and $c$, are determined to 
reproduce the experimental values, $\bar{m}_\pi$ and $\bar{F}_\pi$, 
of $m_\pi$ and $F_\pi$ at $T=0$; 
the resultant values are $m_q=2.26$~MeV and $c=(0.333~{\rm GeV})^3$. 
This parameter set reproduces masses and decay constants of 
$\pi$, $\sigma$, $\rho$, $a_1$ mesons at $T=0$ within $\sim 10 \%$ 
error~\cite{EKSS,GSUY}. 
Furthermore, when $T=0$, the present model satisfies 
the Gell-Mann-Oakes-Renner (GOR) relation 
$\bar{m}_\pi^2 \bar{F}_\pi^2 = 2m_q c$~\cite{EKSS,GMTY}.

In principle, the chiral condensate $c$ depends on $T$, 
but it is known by 
lattice QCD~\cite{Karsch1} and 
the chiral perturbation theory to three loops~\cite{Leutwyler1} 
that $c$ little depends on $T$ in the region of $T < 100$~MeV. 
So we focus our discussion on the temperature region for a while 
and then neglect the weak $T$ dependence of $c$.

Figure~\ref{vector} shows that 
the $T$ dependence of $m_{\pi}$ and $F^t_{\pi}$ is almost constant in the 
region $T < 100$ MeV; 
more precisely, $m_{\pi}$ is slightly reduced and 
$F^t_{\pi}$ is slightly enhanced there. 
This is consistent with the result of 
the chiral perturbation theory to two loops~\cite{Toublan}. 
Furthermore, 
$m_{\sigma}$ is largely reduced as $T$ increases, as shown in 
Fig.~\ref{vector}. This is also consistent 
with the result of lattice OCD~\cite{Karsch2}; 
note that the pole mass is evaluated in the lattice calculation. 
Thus, the present model qualitatively simulates $T$ dependence of 
real QCD in the region $T < 100$ MeV.

In Fig. \ref{vector}, the $T$ dependence of $m_a$ is relatively 
weaker than that of $m_{\rho}$. The difference stems  
from a term $-(g_5v)^2/(zf)^2$
which is present in Eq. (\ref{axial1}) for the axial vector meson 
but not in the corresponding equation (\ref{vectoreom}) 
for the vector meson. The term suppresses partially the enhancement of 
the mass term ${m_a^2/f^4}$ near $T_c$, since the two terms 
have opposite signs to each other. 
In Fig. \ref{vector}, we can also see a tendency for $m_{\pi}$ 
to become small as $T$ reaches $T_c$. 
When $T \approx T_c$, the second term of Eq. (\ref{axial2}) 
is enhanced by the factor $f^{-2}$ compared with 
the first term. This might indicate that $\varphi(z) \approx - \pi(z)$.  
Inserting this relation into Eq. (\ref{axial3}), we can speculate that 
$m_{\pi} \propto vf$ near $T_c$.

\vspace{.3cm}

\begin{figure}[htbp]
\begin{center}
\voffset=15cm
  \includegraphics[width=6cm,height=6cm]{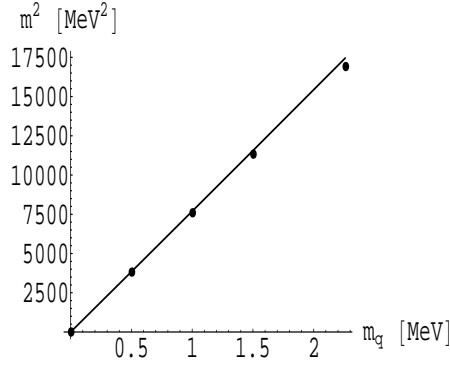}  
\caption{The $m_q$ dependence of $m_{\pi}^2$ at $T=100$~MeV. 
The solid line corresponds to $m_{\pi}^2(T)$ and 
the closed circles to $2m_q c/{F^t_\pi}^2(T)$. 
}
\label{GOR}
\end{center}
\end{figure}

{\bf 3-4.} 
In real QCD, a generalized GOR relation, $m_\pi^2(T){F_\pi^t}^2(T) = 2m_q c$, 
is satisfied 
for finite $T$~\cite{Pisarski,Toublan}. 
Within the framework of the holographic QCD, 
a necessary condition for the GOR relation to be satisfied 
is that Eq. (\ref{axial0}) agrees with 
Eq. (\ref{axial2}) with $\pi(z)=0$~\cite{EKSS}. 
This condition is satisfied for any $T$ in the present model. 
Thus, the present model is expected to reproduce the generalized GOR relation 
properly.

Whether the generalized GOR relation is realized in the present model 
is tested through $m_{\pi}(T)$ and ${F^t_\pi}(T)$ calculated numerially. 
Figure. \ref{GOR} shows the $m_q$ dependence of 
$m_{\pi}^2(T)$ and $2m_q c/{F^t_\pi}^2(T)$ by 
the solid line and the closed circles, respectively. 
Here the case of $T=100$~MeV is taken as an example. 
The $m_{\pi}^2(T)$ shown by the solid line 
tends to zero as $m_q$ deceases, 
as a reflection of the Nambu-Goldstone theorem at finite $T$.
Comparing the solid line and the closed circles, one can see that 
the generalized GOR relation is well satisfied in the present model.

\vspace{.3cm}
{\bf 3-5.} 
The pion velocity $v_{\pi}$ in the thermal midium 
can be estimated from 
$F^t_\pi(T)$ and $F^s_\pi(T)$ as 
${v_{\pi}}^2 = F^s_\pi(T)/F^t_\pi(T)$~\cite{Pisarski,SS}. 
The pion velocity $v_{\pi}$ thus evaluated decreases 
quite slowly from 1 to 0.97 
as $T$ increases from 0 to 100~MeV; 
here we assume $c$ does not depend on $T$.   
Now the value of $v_{\pi}$ is estimated in the limit of $T\to T_c$
by solving the Eqs.(\ref{axial0}) and (\ref{axial2}) at the chiral
limit $m_q=0$. At $T_c$, we expect the restoration of chiral symmetry,
so we set as $v(z)=0$ in the equations and the solutions are given
as 
\beq
 A_i=a_1+a_2 {\tanh^{-1}(z/z_T)^2\over 2z_T^2},
\qquad A_0=b_1+b_2 z^2.
\eeq
The constants $a_{1,2}$ and $b_{1,2}$ are determined by the boundary
conditions, $A_{i,0}|_{z=0}=1$ and 
$\partial_zA_{i,0}|_{z=z_m}=\epsilon$, and we
take the limit $\epsilon\to 0$ after obtaining $v_{\pi}$. Thus we
get
\beq
  v_{\pi}^2=\sqrt{1-({T\over T_c})^4} \propto t^{1/2}~, 
\eeq
where $t=(T_c-T)/T_c$. This indicates $v_{\pi}(T_c)=0$. 
This is consistent with the result 
obtained by Son-Stephanov~\cite{SS}. 
The corresponding critical exponent is $\nu=1/2$ in our case.

\vspace{.3cm}

{\bf 4.} 
The model we present is the first holographic model to describe 
QCD in the region of $T<T_{c}$ nonperturbatively. 
In this model, we take the 5d AdS-Schwarzschild solution 
as the gravity background, and cut off the extra dimension 
at an appropriate infrared point $z_m$ in order to 
introduce the confinement in a gauge theory dual to the gravity. 
The present model then describes the confinement phase in the region $T<T_{c}$.

The present model makes it possible to evaluate 
$T$ dependence of meson masses and decay constants nonperturbatively. 
In the region $T < 100$~MeV, the pion mass and the pion decay 
constant little depend on $T$. This is consistent 
with the result of the chiral perturbation theory to two loops~\cite{Toublan}. 
In contrast, the $\sigma$ meson mass is largely reduced as $T$ increases. 
This is consistent with the result of lattice calculation~\cite{Karsch2}. 
Furthermore, 
the Nambu-Goldstone theorem and the generalized GOR relation are 
assured at any finite $T$.
Thus, the present model reproduces 
the qualitative properties of QCD in the region $T<T_{c}$. 
We then conclude that the present model is reliable enough to make qualitative
predictions on QCD in the region $T<T_{c}$.

Our prediction of the pion velocity is $v_{\pi}(T_c)=0$. 
This agrees with the result of Son-Stephanov~\cite{SS} 
based on scaling and universality. 
Thus, the present result also supports the statement that 
the measured pion velocity 0.65~\cite{Cramer}, deduced from the pion spectra 
observed by STAR~\cite{STAR} at RHIC, would be a signal of QCD phase 
transition.

The present model has three parameters; quark mass $m_q$,  
the chiral condensate $c$ and the cutoff $z_m$. 
In principle, $c$ and  $z_m$ should be determined dynamically and 
be functions of $T$. The $T$ dependence of these quantities 
can not be obtained by the present model. 
The lattice calculation shows that 
the pion mass somewhat increases near $T_{c}$. If it is true, 
this would imply that $z_m$ becomes smaller there, since smaller $z_m$ 
yields larger pion mass. 
Furthermore, it is well known that $c$ has a strong $T$ dependence near $T_c$. 
This is essential for properties of light mesons near $T_c$. 
Thus, it is quite important to 
improve the present model so that $c$ and $z_m$ 
can be determined dynamically. Furthermore, 
 the vector meson mass and decay constant 
do not depend on $c$ in the present model.  
This is also an important issue to be solved in future.

\vspace{.3cm}
K. G thanks Y. Nakano for enlightenment on finite temperature field theory.
This work has been supported in part by the Grants-in-Aid
for Scientific Research (13135223)
of the Ministry of Education, Science, Sports, and Culture of Japan.



\begin{thebibliography}{99}

\bibitem{AdS}
  A.~Karch and E.~Katz, 
  JHEP {\bf 0206}, 043(2003) [hep-th/0205236].
  M.~Kruczenski, D.~Mateos, R.C.~Myers and D.J.~Winters, 
  JHEP {\bf 0307}, 049(2003) [hep-th/0304032].
  M.~Kruczenski, D.~Mateos, R.~C.~Myers and D.~J.~Winters,
  JHEP {\bf 0405}, 041 (2004) [arXiv:hep-th/0311270].
  J.~Babington, J.~Erdmenger, N.~J.~Evans, Z.~Guralnik and I.~Kirsch,
  Phys.\ Rev.\ D {\bf 69}, 066007 (2004) [arXiv:hep-th/0306018].
  N.~J.~Evans and J.~P.~Shock,
  Phys.\ Rev.\ D {\bf 70}, 046002 (2004) [arXiv:hep-th/0403279].
  C.~Nunez, A.~Paredes and A.V.~Ramallo, 
  JHEP {\bf 0312}, 024(2003) [hep-th/0311201]. 
 T. Sakai and S. Sugimoto, Prog.Theor.Phys.113(2005)843-882,
 [hep-th/0412141];  hep-th/0507073. 
  K. Ghoroku and M. Yahiro, 
  Phys.\ Lett.\ B {\bf 604}, 235 (2004) [arXiv:hep-th/0408040].

\bibitem{GSUY} K. Ghoroku, T. Sakaguchi, N. Uekusa and M. Yahiro,
 Phys. Rev. D71(2005)106002, [arXiv:hep-th/0502088].


\bibitem{EKSS}
  J.~Erlich, E.~Katz, D.~T.~Son and M.~A.~Stephanov,
   hep-ph/0501128.
\bibitem{RP}
  L.~Da Rold and A.~Pomarol,
   Nucl.\ Phys.\ B {\bf 721}, 79 (2005) [hep-ph/0501218];
   hep-ph/0510268.
\bibitem{TB}
  G.~F.~de Teramond and S.~J.~Brodsky,
  Phys.\ Rev.\ Lett.\  {\bf 94}, 201601 (2005) [arXiv:hep-th/0501022].

\bibitem{GMTY} K. Ghoroku, N. Maru, M. Tachibana and M. Yahiro,
 hep-ph/0510334.

\bibitem{KLS} E.~Katz, A. Lewandowski and M.D. Schwartz,
  hep-ph/0510388.

\bibitem{STAR}
J. Adams et al. [STAR Collaboration], Phys. Rev. Lett. {\bf 92}, 112301(2004) 
[nucl-ex/0411036].



\bibitem{Karsch1}
  F. Karsch and E. Laermann,
  hep-lat/0305025.

\bibitem{Leutwyler1}
P. Gerber and H. Leutwyler, Nuc. Phys. B321, 387 (1989).


\bibitem{Toublan}
D. Toublan, Phys. Rev. {\bf D56}, 5629(1997) [hep-ph/9706273].

\bibitem{Karsch2}
  F. Karsch, Nucl. Phys. Proc. Suppl. {\bf D83}, 14(2000) [hep-lat/9909006]. 

\bibitem{Pisarski}
R.D. Pisarski and M. Tytgat, Phys. Rev. {\bf D54}, 2989(1996) [hep-ph/9604404].

\bibitem{SS}
D. T. Son and M. A. Stephanov, Phys. Rev. Lett. {\bf 88}, 202302 (2002).


\bibitem{Cramer}
J. G. Cramer, G. A. Miller, J. M. S. Wu and J. H. S. Yoon, Phys. Rev. Lett. 
{\bf 94}, 102302(2005).






\end{thebibliography}
\end{document}